\documentclass{article}    

\usepackage{graphicx}

\title{\textbf{The Case Against \mbox{Objective Measurement}}}  
\author{Richard Mould\footnote{Department of Physics and Astronomy, State University of New York, Stony Brook,
\mbox{New York} 11794-3800}}  
\date{}    
   
\begin{document}             

\maketitle              

\begin{abstract}

  We consider two straightforward rules that govern the stochastic choice in a single quantum mechanical event.  They are
shown to lead to absurd results if an ``objective" state reduction is allowed to compete with an ``observer" state
reduction.  Since the latter collapse is empirically verifiable, the existence of the former is thrown into question.  
   
\end{abstract}

\section*{Introduction}

	Let a particle and detector interact.  The state of the system prior to interaction is given by 

\begin{displaymath}
\Phi(t) = exp(-iHt)\psi_iD_i
\end{displaymath}
where $\psi_i$ is the initial state of the particle, and $D_i$ is the initial state of the detector.  After interaction at
time $t_0$, the particle and detector will become entangled and evolve into two decoherent components
\begin{equation}
\Phi(t \ge t_0) = \psi(t)D_0 + D_1(t)
\end{equation}
where the first is the partially scattered particle together with the detector in its ground state, and the second is the
detector in its excited or capture state.  The second component is zero at $t_0$. 

In the following reducto ad absurdum argument, I first assume that eq.\ 1 may stochastically reduce to either
$\psi(t)D_0$ or $D_1$ without the benefit of an observer.  I will then show that the result is not supportable when an
observer is added to the mix.

\section*{The Rules}

I do not propose a \emph{theory} of collapse.  The rules given below describe the collapse of a wave function in a way that
is independent of theory.  They define the circumstances that trigger a collapse, they specify the timing consistent with
observation and Hamiltonian dynamics, and they do so without having to provide a separate theoretical structure.    A
collapse refers here to an individual event, not an ensemble of events.  So $D_1$ is the sole surviving state in one
collapse, and
$\psi(t)D_0$ is the survivor in another.  Potentially, each of these collapses follows its own rule.  The first rule (A)
governs a collapse to the state $D_1$, and the second \mbox{rule (B)} governs a collapse to the state
$\psi(t)D_0$. 

\vspace{0.3cm}
\textbf{Rule (A)}: \emph{The probability current J flowing from the first to the second component in eq.\ 1 gives the
probability per unit time that there will be a stochastic choice of state $D_1$, causing a collapse of the superposition to
$D_1$.}
\vspace{0.3cm}

The probability current $J$ is equal to the time rate of change of the square modulus.  It is the rate at which the first
component decreases and the second component increases due to current flow from the first to the second.   

The total probability of a collapse to $D_1$ is equal to an integral of $J$ from the beginning of the interaction to the
end.  That amount will depend on the cross section of the interaction and the length of time that the detector is exposed
to the oncoming particle.  This means that the collapse (corresponding to a particle capture) can occur at any time during
the interaction, where both the time of the collapse and the total probability of its happening are determined by the
probability current.  So whatever theory of collapse one might entertain, it must make use of the magnitude and duration
of this current.  Otherwise, the theory will conflict with Hamiltonian dynamics.\footnote{To simplify matters I assume
that a collapse, once begun, is instantaneous.  The following argument does not critically depend on that
assumption.}$^,$\footnote{One thing is certain.  A theory is wrong that requires an (almost) immediate state reduction
because of the macroscopic nature of the detector; for surely the size of the detector has nothing to do with the current
flowing into it. Circumstances might make the current flow in eq.\ 1 indefinitely small; in which case, the macroscopic
superposition might last for an indefinitely long period of time.  This, I believe, is the basic flaw in the
gravitational theory of Penrose \cite{RP} and in the spontaneous reduction theory of Ghirardi, Rimini, and \mbox{Weber
\cite{GRW}}.  Their `timing' is off.  Macroscopic states donÕt always collapse as soon as they are created.}

 Now consider  rule (B) that governs a collapse to $\psi(t)D_0$ when there is no capture in an individual event. 
This one is difficult.  The trouble is that it makes no sense to declare \mbox{`no-capture'} while the interaction is
still in progress.  Doing so would abort \mbox{rule (A)}.  There is no moment of time during the interaction when a
no-capture collapse can occur without prematurely abandoning the possibility of a capture.  Therefore, a collapse to
$\psi(t)D_0$ can only occur when the interaction is complete; that is, assuming that the state is not first collapsed by
rule (A) or by the intervention of an observer.    

\vspace{0.3cm}
\textbf{Rule (B)}: \emph{If the superposition $\psi(t)D_0 + D_1(t)$ survives to the end of an interaction, then it will
collapse to the current contributing component $\psi(t)D_0$, making it the sole survivor.   }
\vspace{0.3cm}

	Rule (B) follows from the necessity to preserve rule (A).  However, there are consequences that make this rule
untenable.  For simplicity, suppose that the approaching particle has a total probability of capture equal to 60\%. 
Let an observer look at the detector before the interaction is complete - at a time when there is only a 50\% probability
of capture.  Rule (A)  then requires that the observer find the detector in its capture state 50\% of the time;  and
otherwise, \mbox{rule (B)} requires that the observer find the superposition of eq.\ 1 since the interaction is
not yet over.  The observer is therefore exposed to a strange combination of pre-quantum and quantum states, consisting of
$D_1$ plus the superposition in \mbox{eq.\ 1}.  This will lead to the following observational absurdity.

The state of the system is listed on the left in fig.\ 1 and the corresponding experience of the observer (following the
Born interpretation) is listed on the right.  From the first row, the observer will find the detector in the state $D_1$ a
full 50\% of the time.  From the second row, $D_1$ will be observed  \emph{half} of 50\% of the time because of the
Born rule applied to the superposition, and $D_0$ will be observed for the \emph{other half} of that time.  All together,
$D_1$ will be observed 75\% of the time, and $D_0$ will be observed 25\% of the time.  Our initial assumption was that the
particle is captured 50\% of the time (up to the time of the observation), so there is something absurdly wrong with this
result?

\begin{figure}[b]
\centering
\includegraphics[scale=0.8]{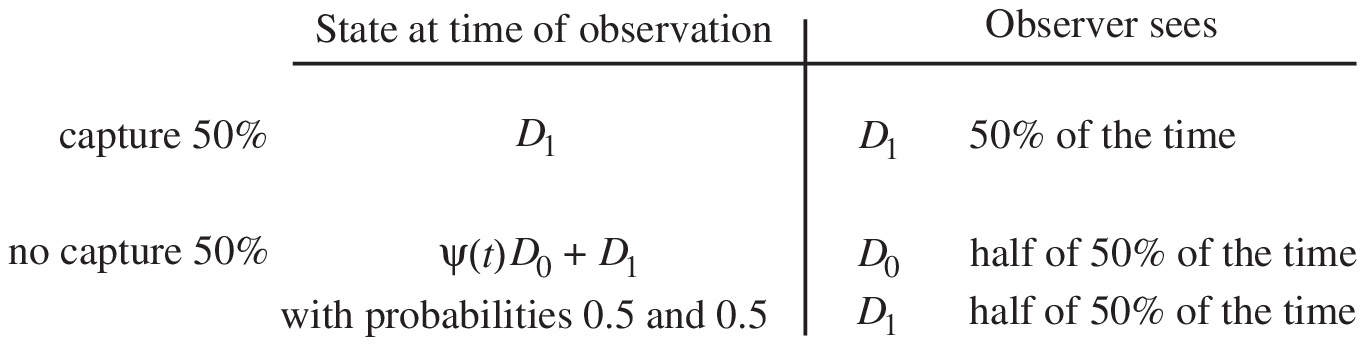}
\center{Figure 1}
\end{figure}

Rule (A) is surely correct if the Hamiltonian is to be believed.  Because of that, a no-capture collapse is unavoidably
delayed until the interaction is terminated is some way, and this sets up the absurd result in fig.\ 1.  The conundrum
posed here is more fundamental than any possible `theory' of measurement.  It has to do with the difficulty in finding a
straightforward empirical formula (i.e., a set of rules A \mbox{and B)} that accurately portrays the `timing' of the two
possible reductions.  Since rules of this kind should apply generally to any interaction, their failure to apply in this
situation suggests that absurdity will follow whenever objective state reduction is allowed to compete with the reduction
associated with an observer.  

The above argument shows that observer state reduction cannot compete with objective state reduction in this
interaction.  Since an observation (resulting in reduction) can surely be arranged, it is strongly suggested that an
objective reduction will not occur in this or in any similar situation.  More generally, it is suggested that there is
\emph{no such thing} as an objective measurement.  Quantum mechanics doesn't need objective measurement, there is no
empirical or presently existing theoretical reason for it, so Occam's razor seems to require that we reject
the idea.  It may only be a remnant of classical thinking.

\section*{Rules Governing Observer Reduction (Observer Measurement)}

On the other hand, it is possible to discover consistent and well-behaved rules that describe \emph{observer reductions}
alone. This has been done in a previous publication \cite{RM}.   These exclusively observer-based rules make it possible
to include the observer in a quantum mechanical system, and they place novel
\mbox{macro-theoretical} constraints on participating brain states.

Standard quantum mechanics with the Born interpretation does not allow the observer to be included in the system.  It
allows an observer to look at the system at one particular time, but it does not let him follow further developments in a
continuous way.  He can come back later and see what has happened, but he cannot stay and become part of the system. This
limitation on the observer is overcome by the proposed observer-based rules \cite{RM2}. 

	There are five of these rules in ref.\ 3, the first four of which give a consistent and accurate empirical description of
observation measurement. The first of these replaces the Born rule.  It replaces the static notion of
\emph{probability} with the dynamic notion of \emph{probability/time}.  It is this, together with the other rules,
that allows the expansion of quantum mechanics to include conscious observers on a continuing basis. 

In addition, these rules make novel synthetic statements about brains.  Two of the rules are \emph{selection rules} that
limit the transitions that can take place between those brain states that are the basis states of state reduction.  As
such, the rules place significant macro-theoretical constrains on the brains of conscious observers.  

The proposed rules in ref.\ 3 are successfully applied to a number of different situations, including two different
versions of the Schr\"{o}dinger cat experiment.  I believe they are generalizable to any interaction that includes an
observer.  Therefore, they are the empirical guidelines that must be met by any theory that claims to explain observer
based state reduction in quantum mechanics, whether or not objective reduction is also validated.

\end{document}